\documentclass[aps,twocolumn,superscriptaddress,showpacs,showkeys,amsmath,amssymb,floatfix]{revtex4}
\usepackage{graphicx}
\usepackage{epsfig}
\usepackage{subfigure}
\usepackage{mathtools}
\usepackage{dcolumn}
\usepackage{bm}
\usepackage{amssymb}
\usepackage{calrsfs}
\DeclareMathAlphabet{\pazocal}{OMS}{zplm}{m}{n}
\usepackage{amsmath}
\usepackage{subfigure}
\newcommand{\be}{\begin{equation}}
\newcommand{\ee}{\end{equation}}
\newcommand{\bea}{\begin{eqnarray}}
\newcommand{\eea}{\end{eqnarray}}

\newcommand{\pro}{\partial}

\newcommand{\ba}{\begin{array}}
\newcommand{\ea}{\end{array}}

\newcommand{\nn}{\nonumber}

\newcommand{\La}{\mathcal{L}}

\newcommand{\Da}{\mathcal{D}}
\newcommand{\Ha}{\mathcal{H}}

\newcommand{\Fa}{\mathcal{F}}

\newcommand{\Ba}{\mathcal{B}}

\begin{document}

\title{Quantum stability of non-linear wave type solutions\\  with intrinsic mass parameter in QCD}
\bigskip
\author{Youngman Kim}
\affiliation{Rare Isotope Science Project, Institute for Basic Science,
Daejeon 305-811, Korea}
\email{ykim@ibs.re.kr}
\author{Bum-Hoon Lee}
\affiliation{Asia Pacific Center of Theoretical Physics,
Pohang 790-330, Korea}
\affiliation{CQUEST, Sogang University, Seoul 121-742, Korea}
\email{bhl@sogang.ac.kr}
\author{D.G. Pak}
\affiliation{Asia Pacific Center of Theoretical Physics,
Pohang, 790-330, Korea}
\affiliation{CQUEST, Sogang University, Seoul 121-742, Korea}
\affiliation{Chern Institute of Mathematics, Nankai University,
Tianjin 300071, China}
\email{dmipak@gmail.com}
\author{Chanyong Park}
\affiliation{Asia Pacific Center of Theoretical Physics,
Pohang, 790-330, Korea}
\affiliation{Department of Physics, Postech, Pohang 790-784, Korea.}
\email{chanyong.park@apctp.org}
\author{Takuya Tsukioka}
\affiliation{School of Education, Bukkyo University, Kyoto 603-8301, Japan}
\email{tsukioka@bukkyo-u.ac.jp}

\begin{abstract}
The problem of existence of a stable vacuum field 
in a pure quantum chromodynamics (QCD) is revised. Our approach is based
on using classical stationary non-linear wave type solutions with intrinsic
mass scale parameter. Such solutions can be treated as quantum mechanical
wave functions describing massive spinless states in quantum theory.
We verify whether non-linear wave type solutions can form a stable 
vacuum field background within the framework of effective action formalism.
We demonstrate that there is a special class of stationary generalized Wu-Yang 
monopole solutions which are stable against quantum gluon fluctuations.
\end{abstract}
\maketitle

\section{Introduction}

The origin of the quark/color confinement and mass gap in quantum chromodynamics
represents the most principal problem in foundations of the theory of strong interactions
\cite{colorconft}. One of the most attractive mechanisms of the quark confinement
is based on the dual Meissner effect in color superconductor by means of monopole
condensation \cite{nambu74,mandelstam76,polyakov77,thooft81}.
If such a stable monopole condensate is generated, it
will immediately imply the confinement \cite{ezawa82,suzuki80,suganuma95}
which has been confirmed in lattice simulations
\cite{kronfeld87,suzuki90,stack94,shiba94,bali96}.
Theoretical foundation of the confinement mechanism with the
dual Meissner effect encounters several obstacles. Among them,
the realization of physical monopole solutions in the standard
QCD and quantum stability of monopole condensation
represent a long-standing problem since late 1970s when the
Savvidy-Nielsen-Olesen vacuum instability was found \cite{savv, N-O}.
So far, neither a regular monopole solution
nor a strict construction of a stable color magnetic condensate has been known
in the framework of the basic standard theory of QCD.
This causes serious doubts that the known Copenhagen ``spaghetti'' vacuum
and other models of QCD vacuum can provide rigorous microscopic description of the
vacuum structure
\cite{niel-nino,niel-oles,amb-oles1,amb-oles2,bordag,pak05}.

In the present paper we elaborate an idea that classical stationary non-linear
wave type solutions can be treated in a quantum mechanical sense
and describe physical states in quantum theory.
The idea that stationary non-solitonic wave solutions correspond to particles or quasi-particles
was sounded long time ago \cite{derr, jackiw77,jackiwRMP}.
Our goal is to find a proper regular stationary solution which will be
stable against quantum gluon fluctuations within the formalism of the
effective action in one-loop aproximation. Such a stable field configuration
can serve as a structure element in further construction of 
a true QCD vacuum.
There is a wide class of known stationary non-linear wave solutions
\cite{mat1,lahno95,smilga,frasca09,tsap, p1}
which possess non-trivial features: the presence of mass scale parameters,
non-vanishing longitudinal components of color fields along
the propagation direction,  color magnetic charge and vanishing
classical spin density operator.
This gives a hint that some of such classical solutions 
describe quantum states corresponding to
massive spinless quasi-particles which might
lead to formation of a stable vacuum condensate. 
Surprisingly, we show that there is a special class
of stationary spherically symmetric monopole solutions which
possesses quantum stability.

The paper is organized as follows;
in Section II we overview the main critical points in
the vacuum stability problem and outline possible ways towards
construction of a stable vacuum field configuration.
Quantum stability of non-linear plane wave solutions is considered
in Section III.
A careful analysis shows that in spite of several attractive properties of such solutions
the non-linear plane waves are unstable against vacuum gluon fluctuations.
In Section IV we consider quantum stability of a recently proposed stationary monopole 
solution \cite{p1}
which represents a system of a static Wu-Yang monopole interacting to
off-diagonal components of the gluon field.
We have proved that such a generalized monopole solution
provides a stable vacuum field background in the effective action of QCD
in one-loop approximation.
Conclusions and discussion of our results are presented in the last section.
An additional qualitative analysis of quantum stability of the 
stationary monopole field is given in Appendix. 

\section{Vacuum stability problem}

Let us consider the structure of the QCD effective action
in the presence of constant homogeneous classical fields and
expose the critical issues of vacuum instability for that simple case.
In order to study the vacuum structure in quantum field theory
it is suitable to apply
a quantization scheme based on the functional integral formalism and calculate
the quantum effective action with a properly chosen classical background field.
The background field satisfying the classical equations of motion corresponds to
a vacuum averaged value of the quantum field operator in the presence of a source
or in the adiabatic limit when the external source vanishes at
time $t\to +\infty$.
A non-trivial vacuum structure can be retrieved from the behavior of the
effective potential and from the structure of the effective action.
In general the effective potential admits several local minima,
and only the lowest and stable one determines a true physical vacuum.
Moreover, the symmetry properties of the vacuum state determine
fundamental properties of the theory such as the type of symmetry breaking,
possible phase transitions, etc. The knowledge of
the analytic structure of the effective action represents
an important step which verifies whether a non-trivial classical
vacuum in the theory corresponds to
a physical vacuum at quantum level.
As usual, the presence of an imaginary part of the effective action
indicates  vacuum instability.

   We concentrate mainly on the structure of the effective action in the
case of a pure $SU(2)$ QCD. For the case of constant homogeneous classical
background field the effective action can be calculated in a complete form
in one loop approximation. We start with a classical Lagrangian
of Yang-Mills theory
\bea
{\La}_0&=&-\dfrac{1}{4} F_{\mu\nu}^aF^{\mu\nu}_a,
\label{classL}
\eea
with $$
F_{\mu\nu}^a= \pro_\mu A_\nu^a-\pro_\nu A_\mu^a
+g\epsilon^{abc}A_\mu^b A_\nu^c.
$$
The space-time indices $\mu, \nu$ and those for colors $a, b, c$
run through $0, 1, 2, 3$ and $1, 2, 3$, respectively.
We work with the convention $g_{\mu\nu}={\rm diag}(-1, 1, 1, 1)$ and
$\epsilon^{123}=1$.

An initial gauge potential $A_\mu^a$ is split into a classical, $\Ba_\mu^a$,
and a quantum, $Q_\mu^a$, parts
\bea
A_\mu^a=\Ba_\mu^a+Q_\mu^a.\label{splitpot}
\eea
One should stress that the classical gauge potential $\Ba_\mu^a$ must
be a solution to classical Euler-Lagrange equations of motion.
Only in that case the external classical field $\Ba_\mu^a$ can be treated
as a vacuum averaged value of the quantum operator $A_\mu^a$
in a consistent manner with the effective action formalism.
One should note that a static homogeneous classical gauge potential $\Ba_\mu^a$
can not provide a constant field strength unless the gauge symmetry is broken.
The field $\Ba_\mu^a$ is defined as a vacuum averaged value of the
quantum operator $A_\mu^a$
in the limit of vanishing source $J_\mu^a \to 0$ during the time evolution
($t\to +\infty$)
\bea
\Ba_\mu^a &=& \langle 0|A_\mu^a|0\rangle_{J\to 0},
\eea
where $|0\rangle$ is a vacuum state.
It is clear that due to the gauge and Lorentz invariances
the vacuum averaged value of the gauge potential must be identically zero,
i.e.\,  $\Ba_\mu^a \equiv 0$.
A partial solution to this problem was suggested
by proposing the ``spaghetti'' vacuum model where
the vacuum is represented
by a statistical ensemble of vortex domains
which leads to a zero mean value of the gauge field.
However, in such cases one encounters two principal obstacles:
(i) the statistical field ensemble does not represent
an exact solution to the classical equations of motion,
and (ii) at the microscopic scale each domain or a single vortex
causes instability due to non-vanishing contribution
to the imaginary part of the effective action.
So that a statistical ensemble does not provide a
microscopic theory of the vacuum structure on a firm basis of the
standard quantum field theory.

With these preliminaries let us write down the main equations
which allow to retrieve the analytic structure of the effective action
for arbitrary background gauge field configuration. It is convenient to
choose a covariant Lorenz gauge fixing condition  for the quantum gauge potential
\bea
{(\Da}_\mu Q^{\mu})^a=0,
\eea
where ${\Da}_\mu^{ab}= \delta^{ab}\partial_\mu+g\epsilon^{acb}\Ba_{\mu c}$
is a covariant derivative
including the background gauge field potential $\Ba_\mu^a$.
Applying a standard functional technique, one can express the
one-loop correction to the classical action
in terms of functional determinants
\bea
S^{\rm 1\, loop}_{\rm eff}&=&-\dfrac{i}{2} \ln {\rm Det} [K_{\mu\nu}^{ab}]+i
\ln {\rm Det} [M_{\rm FP}^{ab}], \label{lndet} \\
%
%
K_{\mu\nu}^{ab}&=&
-g_{\mu\nu}({\Da}^\rho {\Da}_\rho)^{ab} -2 \epsilon^{acb}{\Fa}_{\mu\nu}^c, \nn \\
M_{\rm FP}^{ab}&=&-({\Da}^\rho{\Da}_\rho)^{ab}, \nn
\eea
where ${\Fa}_{\mu\nu}^a$ is a background field strength and the
operators $K_{\mu\nu}^{ab},~ M_{\rm FP}^{ab}$ correspond to
one-loop contributions of
gluons and Faddeev-Popov ghosts.
One should stress that expression (\ref{lndet}) represents an exact one-loop result for
arbitrary configuration of the background gauge field $\Ba_\mu^a$.
One can obtain similar expressions for the one-loop functional determinants
in the case of using an initial temporal gauge for the quantum
gauge potential and an additional Coulomb type gauge condition which
fixes the residual symmetry.

\subsection{A constant Abelian magnetic field}

Let us consider first a simple case of the Savvidy vacuum \cite{savv}
based on a classical solution for the constant homogeneous magnetic field of Abelian type
defined by the gauge potential
${\Ba}_\mu^a=g_{\mu 2}\delta^{a3} x H$. The gauge field strength ${\Fa}_{\mu\nu}^a$
has only one non-vanishing magnetic component ${\Fa}_{12}^3=H$.
In that case the expression for the one-loop correction to the  effective action (\ref{lndet})
can be simplified to a form
\bea
S_{\rm eff}^{\rm 1\, loop}&=&
i\!\sum_{S_z=\pm1} 2\, {\rm Tr} \ln [-{\Da}^\mu{\Da}_\mu+2 g H S_z],
\label{Sdet}
\eea
where $S_z=\pm 1$ is a spin projection onto the $z$-axis of the gluon
which is treated as a massless vector particle in the Nielsen-Olesen approach \cite{N-O}.
It is clear that the operator inside the logarithmic function is not positively defined
for $S_z=-1$. This causes an imaginary part of the effective action
and implies the Nielsen-Olesen unstable ``tachyon'' mode \cite{N-O}.
An important issue is that the origin of the vacuum instability
is due to a specific interaction structure of the non-Abelian gauge theory;
namely, due to the anomalous magnetic moment interaction of the vector gluon
 with the magnetic field $H$. Note that the contribution of the Faddeev-Popov ghosts
does not induce any imaginary part since the interaction of spin zero ghost fields with the magnetic field
has no such an anomalous magnetic moment interaction.
The functional determinants in (\ref{Sdet}) can be calculated
using the Schwinger proper time method. With this the effective Lagrangian
can be expressed in a compact integral form
\cite{yildiz80,claudson80,adler81,dittrich83,flory83,blau91,reuter97,chopakPRD}
\bea
{\La}_{\rm eff}^{\rm 1\, loop}\!\!\!
&=&\!\!\! \dfrac{1}{16 \pi^2} \!\!\int_0^\infty \!\!\!\!\!\dfrac{{\rm d}s}{s^{(2-\varepsilon)}}
\dfrac{g H/\mu^2}{\sinh (gHs/\mu^2)} \big (\,
{\rm e}^{-\frac{2gHs}{\mu^2}}+{\rm e}^{\frac{2gHs}{\mu^2}}\big), \nn \\
\eea
where $\varepsilon$ is the ultra-violet cut-off parameter and $\mu^2$ is a mass scale
parameter corresponding to the subtraction point.
The second exponential term in the last equation leads to
a severe infra-red divergence which is reflection of the same anomalous magnetic
moment interaction term in (\ref{Sdet}). 
One can perform an infra-red regularization by 
changing the proper time variable to a pure imaginary one, $s \rightarrow i t$,
\cite{chopakPRD}
\bea
{\La}_{\rm eff}^{\rm 1\, loop}\!\!\!
&=&- \dfrac{1}{8 \pi^2} \!\!\int_0^\infty \!\!\!\!\!\dfrac{{\rm d}t}{t^{(2-\varepsilon)}}
\dfrac{g H/\mu^2}{\sin (gHt/\mu^2)} 
\cos (2gHt/\mu^2). \nn \\
\label{intreg}
\eea
This removes the infra-red divergence, but now one encounters an ambiguity
in choosing contours of the integral due to appearance 
of infinite number of poles at $t=\pi k \mu^2/gH$, $(k=0,1,2,...)$.
We define the integration path $t=0-i \delta$ 
with an infinitesimal number factor $\delta$. 
One can verify that a total residue contribution from  the poles 
reproduces exactly the Nielsen-Olesen imaginary part of the effective Lagrangian \cite{N-O}
\bea
\rm{Im} {\La}&=& \dfrac{1}{8 \pi} g^2 H^2.
\eea
Note that a color electric field causes the vacuum instability due to the
Schwinger's mechanism of charged particle-antiparticle pair creation in the
external electric field. Moreover, in a pure gluodynamics it has been shown that a
homogeneous chromoelectric field $E$ leads to a
negative imaginary part of the effective one-loop Lagrangian \cite{schan82}
\bea
\rm{Im} {\La}&=&- \dfrac{11}{96 \pi} g^2 E^2.
\eea

One concludes that a constant homogeneous color magnetic and electric
field of Abelian type is unstable.
A physical meaning of such  instability is the gluon pair creation in the chromomagnetic field
and the gluon pair annihilation in the case of the chromoelectric background field \cite{schan82}.

\subsection{Non-Abelian constant field configuration}

It has been established that $SU(N)$ Yang-Mills theory admits two types
of constant homogeneous field configurations \cite{leutwyler}.
The first type is represented by  Abelian type gauge potentials which correspond
to the Cartan subalgebra of the Lie algebra $\mathfrak{su}(N)$.
The constant homogeneous fields of the second type originate
from the non-Abelian structure of the gauge field strength due to non-commutativity
of the Lie algebra valued gauge potentials \cite{leutwyler}
\be
\vec F_{\mu\nu}= \vec A_\mu \times \vec A_\nu.
\label{AxA}
\ee
Contrary to the case of Abelian constant color magnetic fields, the non-Abelian
magnetic field admits a spherically symmetric configuration.
It was observed that symmetrization of the Hamiltonian of QCD might help
to cure  the Nielsen-Olesen instability \cite{ragiadakos}.
After the discovery of Savvidy-Nielsen-Olesen
vacuum instability, some attempts have been undertaken to construct a stable vacuum
made of constant non-abelian gauge fields.
The results of studies of such a vacuum lead
to the vacuum instability due to the same origin, i.e.\, the presence of
the anomalous magnetic moment interaction \cite{parth,huang}.

Let us overview shortly the known results with
a purpose to find out a way towards resolving the problem of
vacuum stability.
We consider the following isotropic homogeneous field configuration
of non-Abelian type defined by the classical gauge potential
\bea
\Ba_0^a&=&0, ~~~~\Ba_m^a=\phi(t) \delta_m^a. \label{Bphi}
\eea
Throughout this paper,
we use Latin indices $m, n$ as those for the space components of the
four vectors.
The function $\phi(t)$ may have time dependence to include the case
with  non-vanishing constant color electric field as well.

We will find eigenvalues of the operators $K_{\mu\nu}^{ab},~M^{ab}_{\rm FP}$ in
the weak field approximation
assuming that $\phi(t)$ is a slowly varying function.
In the momentum space representation one has
\renewcommand{\arraystretch}{1.4}
\be
\begin{array}{rcl}
K_{mn}^{ab}&=&\delta_{mn}\big (\delta^{ab} (k^2+2 \phi^2)-2 i \phi \epsilon^{acb}k_c\big) \\
&&-2 \phi^2 (\delta_m^b \delta_n^a-\delta_m^a \delta_n^b), \\
K_{0n}^{ab}&=&2 \epsilon^{ab}{}_n b, \\
M_{\rm FP}^{ab}&=&\delta^{ab} (k^2+2 \phi^2)-2 i \phi
 \epsilon^{acb}k_c=-K_{00}^{ab} ,
\end{array}
\label{funcdet}
\ee
where the time derivative term $b\equiv\partial_0 \phi$ corresponds to components of
a color electric field in the temporal gauge $\Ba_0^a=0$.
In the weak field approximation the fields $\phi$ and $b$ are treated
as constant fields.
To find the eigenvalues of the operators
$K^{ab}_{\mu\nu},\ M_{\rm FP}^{ab}$, let us first calculate
the corresponding matrix determinants with respect to Lorentz and color indices.
After some calculations one obtains
\be
\begin{array}{rcl}
\det K_{\mu\nu}^{ab}&=&L_1 L_2 L_3 L_4, \\
\det M_{\rm FP}^{ab}&=& (2 \phi^2 + k^2) ((2 \phi^2 + k^2)^2 - 4 \phi^2 \vec k^2),
\end{array}
\ee
with
\bea
L_1&=&k^4 - 4 \phi^2 \vec k^2, \nn \\
L_2&=&(2 \phi^2 + k^2)\Big(k^2 (4 \phi^2 + k^2) (6 \phi^2 + k^2) \nn \\
   && -4 \phi^2 (2 \phi^2 + k^2) \vec k^2\Big) + 8 k^2 (6 \phi^2 + k^2) b^2, \nn \\
L_3&=&2 \phi^2 (k^2 - 2 \phi |\vec k|) (3 k^2 - 2 \phi |\vec k|) +
8\phi^4(k^2- \phi|\vec k|)\nn \\
&&
+
k^2(k^2-2 \phi |\vec k|)^2+8(k^2-\phi|\vec k|) b^2, \nn \\
L_4&=&2 \phi^2 (k^2 +2 \phi |\vec k|) (3 k^2+2 \phi |\vec k|) +
8\phi^4(k^2+\phi|\vec k|) \nn \\
&&
+
k^2(k^2+2 \phi |\vec k|)^2+8(k^2+\phi|\vec k|) b^2.\nn
\eea
In the particular case with a constant pure magnetic field background, $b=0$,
our result reduces exactly to the known expressions obtained earlier in \cite{parth},
where it has been shown that all eigenvalues corresponding to the operators
$L_i$ are real.
Explicit expressions for all twelve eigenvalues of the operators $L_i$
in the case of pure magnetic background field $b=0$ were
obtained in \cite{parth}.

The operator  $L_1$ is decomposed into the product of two eigenvalues
\bea
\lambda_{1,2}&=& k^2\pm 2 \phi |\vec k| . \label{unstmod2}
\eea
It is easy to verify that the expression for the $L_2$ is non-negative for
any values of $\phi, b, \vec k$ and $k$.
The operator $L_{3}$ has one real and two complex eigenvalues, and $L_4$ has eigenvalues
which are complex conjugate to the eigenvalues of the operator $L_3$.
In general the complex and negative eigenvalues
of the operators $L_1, L_3$ and $L_4$ cause vacuum instability.

One may observe that Eq. (\ref{unstmod2}) implies negative eigenvalues
for small momentum $\vec k$ of the virtual gluon inside the loop.
Remind that the Nielsen-Olesen unstable mode
originates from the anomalous magnetic moment interaction term
$gHS_z$ in (\ref{Sdet}) which does not depend on the
internal momentum $\vec k$. So, in the case of symmetric field configuration
one has no instability in the limit of zero momentum $\vec k$.
So, the symmetric non-Abelian magnetic field configuration makes the
instability problem more soft,
even though the source of appearance of the negative eigenvalues
remains the same as for the Nielsen-Olesen unstable mode.

The presence of instability of the vacuum made from the non-Abelian gauge field
is somewhat puzzling since one expects that the dynamics
of non-Abelian gauge field should provide a consistent quantum vacuum
in a pure QCD. In this connection one should observe one essential weak point
in the above consideration: the constant non-Abelian gauge field does not
represent a classical solution. Due to this the standard method based on the formalism
of functional integration can not be applied self-consistently to derivation of the one-loop effective action.
This raises a question of whether non-Abelian type magnetic field
can be realized as a strict solution, and, if so, whether such a solution
can provide a stable vacuum.
Note that to find a stable physical vacuum
 one should go beyond one-loop approximation
since at one loop level a quartic self-interaction term in the
initial Yang-Mills Lagrangian is omitted and does not affect a final result.
However, the confinement phenomenon is certainly provided by
self-interaction of gluons. So, the quartic interaction term should be
important as an essential part of non-perturbative dynamics.
The evaluation of an exact two-loop effective action
in QCD represents a hard unresolved problem. To go beyond one-loop approximation
one can implement non-perturbative effects in the structure of the classical
solution used as a background field in the effective action.
We conclude that one should look for a proper non-perturbative and
essentially non-Abelian solution of the classical equations of motion which
can lead to a consistent description of the stable vacuum.

\section{Quantum instability of non-linear plane waves}

       Stationary non-linear wave type solutions can be treated
as quantum mechanical wave functions which describe possible
states in quantum theory. In particular, we are interested in
such a classical solution which are stable against quantum gluon fluctuations.
A known class of non-linear plane wave solutions with a mass scale and zero spin
\cite{mat1,lahno95,smilga,frasca09,tsap,p1} is of primary interest
in our search of possible stable vacuum fields since one expects that a system
of massive spinless particles can form a stable condensate in the classical theory.
The presence of spinless states can help in removing the Nielsen-Olesen instability.
We consider a special plane wave solution in  $SU(2)$ Yang-Mills theory
which possesses a spherically symmetric configuration
in the rest frame \cite{mat1,lahno95,smilga,frasca09,tsap,p1}.
A simple ansatz for non-vanishing components of the gauge potential
reads
\bea
{\Ba}_m^a&=&\delta_m^a \phi(u), \label{ans1}
\eea
where $u\equiv k_0 t$.
Substituting the ansatz into the Yang-Mills equations, we obtain
 an ordinary differential equation
\bea
&& k_0^2 \dfrac{{\rm d}^2\phi}{{\rm d}u^2}+2 g^2 \phi^3=0.  \label{eq1}
\eea
One has the following non-vanishing components
for the color electric and magnetic fields
\be
\begin{array}{rcl}
F^1_{10}&=&F^2_{20}=F^3_{30}=-\pro_t \phi, \\
F_{mn}^a&=&g\epsilon_{mn}{}^{a}  \phi^2.
\end{array}
\label{fieldstr}
\ee
The solution to the equation (\ref{eq1}) is given by the Jacobi elliptic
function
\bea
\phi(u)&=&\dfrac{M}{g} \, {\rm sn} [M t,-1],\label{phi0}
\eea
which is a double periodic analytic function with a periodicity
$T_0=4 K[-1]\simeq 5.244 ... $,  ($M=1$), and $K[-1]$
is a complete elliptic integral of the first kind.
The solution contains a mass scale parameter $M$ due to
the conformal invariance of the equations of motion.

The one-loop effective potential in a constant color electric and magnetic field
possesses a local minimum for a non-zero value of the magnetic field
and for the vanishing electric field. The presence of the electric field in the solution
(\ref{fieldstr}) can lead to instability of the vacuum due to the Schwinger pair creation effect.
However, since the electric field of the solution is represented by a periodic function,
the time dependence may change the stability properties of the vacuum field.
Another advantage of treating the stationary plane wave solutions as a quantum mechanical wave
function describing the vacuum state is that the time averaging leads  naturally
to vanishing of the vacuum expectation value of the gauge potential,
$\langle0|A_\mu^a|0\rangle=0$, whereas the averaged magnetic field remains non-zero.

Now we can study the structure of the functional determinants
in (\ref{funcdet}). It turns out that the matrix operator $K_{mn}^{ab}$
gains complex eigenvalues. The presence of complex eigenvalues makes the analysis
of the structure of the effective action complicate since
in that case one needs to know the analytic structure of
the full effective action in the presence of color magnetic and electric fields
\cite{chopakTMU}. Due to this we consider the structure of the one-loop
effective action in the temporal gauge, $Q_0^a=0$ (the background
gauge field satisfies the temporal condition due to the structure of the ansatz (\ref{ans1})), which simplifies
significantly the analysis of possible unstable modes.
In the temporal gauge one has a known residual gauge symmetry
under the gauge transformations only
with  space dependent gauge parameters. To fix this symmetry one
can impose an additional Coulomb constraint,
$\pro^m Q_{m}^a=0$. Therefore, the calculation of the Faddeev-Popov ghost
determinant becomes more difficult since one should introduce secondary
ghosts. However,
since all ghost fields correspond to interaction of spinless particles
with the magnetic field, they do not cause vacuum instability, and we do
not need to calculate ghost contributions in studying the imaginary part of the
effective action.
With this one can perform functional integration
over the quantum field $Q_\mu^a$ and obtain
the following expression for the matrix operator $K_{mn}^{ab}$
\bea
K_{mn}^{ab}&=&\delta_{mn}\delta^{ab}
\big(\pro^2_{t}-\pro_l^2+2 g^2\phi^2(t)\big)+
2\epsilon^{ab}{}_c {\Fa}_{mn}^c \nn \\
&&
-g \phi(t) \big(
\epsilon^{ab}{}_m\pro_n+\epsilon^{ab}{}_n\pro_m+2
\epsilon^{acb}\delta_{mn}\pro_c
\big). \qquad
\label{Koperator}
\eea
Since the  field $\phi(t)$ does not depend on space coordinates,
one can easily perform Fourier transformation with respect to the space coordinates.
After performing the Wick rotation $t \rightarrow i \tau$,
 one arrives at the following expression for the operator
$K_{mn}^{ab}$ in the momentum space representation
\bea
K_{mn}^{ab}&=&
\delta_{mn}\delta^{ab}(-\pro_\tau^2)+
\delta^{ab} \delta_{mn} (\vec k^2+2 \phi^2)\nn \\
&& +\epsilon^{ab}{}_c (2 \phi^2 \epsilon^c_{~mn}+
2 \phi \delta_{mn} i \vec k^c-\phi\delta_{m}^c i \vec k_n+\phi \delta_n^ci \vec k_m) \nn \\
&\equiv& \delta_{mn}\delta^{ab}(-\pro_\tau^2)+\hat K^{ab}_{mn}.
\eea

One can find the eigenvalues $\hat L_i$  of the matrix operator $\hat K_{mn}^{ab}$
since the field $\phi$ does not depend on the space components of the momentum
\bea
\hat L_1&=&\vec k^2 \nn\\
\hat L_{2,3}&=&\vec k^2+4 \phi^2\pm \phi |\vec k| \nn \\
\hat L_{4,5}&=&\vec k^2+5 \phi^2 \pm \sqrt {\phi^4+6 \phi^2 \vec k^2}, \\
\hat L_{6,7}&=&\vec k^2 \pm \phi |\vec k|, \nn \\
\hat L_{8,9}&=&\vec k^2 \pm 2 \phi |\vec k|. \nn
\eea
With this one has finally nine ordinary second order differential
equations for eigenfunctions of the initial kinetic operator $K_{mn}^{ab}$
\bea
&&(-\dfrac{{\rm d}^2}{{\rm d}\tau^2} +\hat L_{q}) \psi_q=
\lambda_q\psi_q, \quad (q=1, 2, \cdots, 9).\quad   \label{eqL}
\eea

The differential equations containing the operators $\hat L_q$, $(q=6, 7, 8, 9)$
might have negative eigenvalues since the respective
differential operators are not
positively defined at small momenta $\vec k$.
Let us rewrite the equation (\ref{eqL}) in the
case $q=6,7,8,9$ in the following form
\bea
&&-\dfrac{{\rm d}^2 \psi}{{\rm d}\tau^2} +(k^2+ \alpha k \phi(\tau)) \psi=
\lambda \psi,   \label{eq13}
\eea
where $k \equiv |\vec k|$ and $\alpha=\pm 1, \pm 2$.
Note that the classical solution $\phi(\tau)$ is identical to the
original solution $\phi(t)$ in (\ref{phi0}),
since by definition the classical field $\Ba_m^a$ corresponds to a vacuum averaged value of
the quantum operator $A_\mu^a$ in the real Minkowski space-time.
We remind that Wick rotation $t\rightarrow i\tau$
provides a causal structure of the Green function, and it does
not mean that one should treat the classical field $\Ba_m^a$
as a solution of the equations of motion in the Euclidean space-time.

The equation (\ref{eq13}) includes the momentum $k$ as a free positive
parameter, and the quantum vacuum stability of the classical solution
will occur if all eigenvalues of the Eq.(\ref{eq13}) are
non-negative for all values of ``$k$'' and for $\alpha=\pm 1,\pm 2$.
It is convenient to rewrite the equation  (\ref{eq13})  as follows
\bea
&&-\dfrac{{\rm d}^2 \psi}{{\rm d}\tau^2} +V_0 (1-\phi(\tau)) \psi= E \psi,
\label{eq13.2}
\eea
with $ V_0\equiv \alpha k, ~~E \equiv \lambda-k^2+\alpha k $.
The equation represents a Schr\"{o}dinger type
equation for a quantum mechanical problem in
one dimensional space parametrized by
$\tau \geq 0$, and $\psi(\tau)$ is a wave function describing
quantum fluctuations of the virtual gluon. One can make another analogy
that
the equation (\ref{eq13.2}) describes behavior of the electron
in the one-dimensional crystal with a periodic potential.
It is known that such an electron in the crystal
is not localized and can move freely in the whole crystal volume.
The electron wave function is expressed by the periodic Bloch function
and the energy spectrum forms a band structure (see, for ex., \cite{landau9}).
To check whether the equation (\ref{eq13.2}) has negative eigenvalues
it is enough to estimate a lowest energy bound in the first energy band.
For qualitative estimation
we consider first a Schr\"{o}dinger equation with a periodic rectangular potential
\bea
V(\tau)&=&\left \{
\ba{lcll}
+1, &\quad& nT\leq \tau \leq (2n+1)\dfrac{T}{2},  \\
-1,  &\quad& (2n-1)\dfrac{T}{2}\leq \tau \leq n T,
\ea
\right .
\eea
where $n=0,\pm 1, \pm 2, \cdots$.
Analityc expressions for a solution of the Schr\"{o}dinger equation
with the potential $V(\tau)$ and the dispersion relation
can be obtained by solving the equation on a finite
interval $(0\leq \tau \leq T)$ \cite{landau9}.
Taking the shift in the potential height into account and setting $T=1$ one
can find an eigenvalue corresponding to the lowest energy level in the first band
which turns out to be negative,  $E_{\rm lowest}\simeq -0.04$.

The numeric analysis of  Eq.(\ref{eq13.2})
 shows that for $\alpha=1$ there is no negative eigenvalues
for any momentum $k$, and the eigenvalue $\lambda$ approaches zero
from the positive values when $k \rightarrow 0$.
For the case $\alpha=\pm 2$ the numeric solutions of the equation  (\ref{eq13.2})
implies negative eigenvalues for the momentum $k$ in the range $(0\leq k \leq 0.733)$
with the lowest eigenvalue $\lambda_{\rm lowest}=-0.0361$
at $k_0=0.482$.
Note that the scale parameter $M$ in the non-linear plane wave
solution $\phi(x)=M{\rm sn}[Mx,-1]$
leads to rescaling of the eigenvalue $\lambda$ and
does not affect the stability properties
as it should be due to conformal invariance of the original classical Yang-Mills theory.
We conclude that despite several attractive properties the non-linear
plane wave solutions can not provide a stable vacuum field configuration.

\section{A stable spherically symetric monopole field background}

Let us first describe the main properties of the stationary spherically symmetric monopole
solution \cite{p1}.
Due to conformal invariance of the Yang-Mills theory the static soliton solutions
do not exist in agreement with the known Derrick's theorem.
It is somewhat unexpected  that a pure QCD
admits a regular stationary monopole like solution \cite{p1}.
The solution is described by a simple ansatz
which generalizes the static Wu-Yang monopole solution
(in spherical coordinates $(r,\theta, \varphi)$)
\be
A_\varphi^1 =-\psi(r,t) \sin \theta, \quad
A_\theta^2 =\psi (r,t), \quad
A_\varphi^3 = \dfrac{1}{g} \cos \theta, \label{ans2}
\ee
where  $\psi(r,t)$ is an arbitrary function and all
other components of the gauge potential vanish.
In the case of  $\psi(r,t)=0$ the ansatz describes
a Wu-Yang monopole solution which is singular at the origin
$r=0$.  The case $\psi(r,t) =1$ corresponds to a pure gauge field configuration.
For a non-trivial function $\psi(r,t)$
the ansatz (\ref{ans2}) describes a system of a static
Wu-Yang monopole dressed in off-diagonal gluon field.
Substituting the ansatz into the equations of motion,
we obtain a single partial differential equation
\bea
\partial^2_t\psi-\pro^2_{r}\psi+\dfrac{1}{r^2}\psi (g^2 \psi^2-1)=0.    \label{eqpsi}
\eea
The equation (\ref{eqpsi}) was obtained in past by using a spherically-symmetric
 ``hedgehog'' ansatz describing
generalized $SU(2)$ Wu-Yang monopole field configurations
($a=1,2,3$)
\bea
A_m^a= -\epsilon^{abc}\hat n^b \pro_m \hat n^c \Big (\dfrac{1}{g}-\psi(t,r) \Big), \label{hedgehog}
\eea
where $\hat n=\vec r/r$ \cite{p14,p15,p16,p17,p18}. 
Note, the ``hedgehog'' ansatz  (\ref{hedgehog}) is related to the ansatz  (\ref{ans2})
by an appropriate singular gauge transformation \cite{choprd80}.  
We prefer to use the ansatz  (\ref{ans2}) in the so-called Abelian
gauge \cite{choprd80} since such a representation allows to inteprete
the our monopole solution as a static Wu-Yang monopole
interacting to dynamic off-diagonal gluons presented by the filed $\psi(r,t)$. 
Note, that the ansatz in the Abelian gauge admits
generalization to the case of $SU(N)$ stationary Wu-Yang monopole
solutions and it is suitable for description of a stationary system
of monopoles and antimonopoles located at different points. 

It had been shown that the equation (\ref{eqpsi}) admits a wide class of 
time dependent solutions including
non-stationary solitonic propagating solutions in the effective two-dimensional space-time 
$(r,t)$ \cite{p14,p15,p16,p17,p18}. Surprisingly, a stationary regular Wu-Yang type monopole solution with a finite energy density everywhere was missed in previous studies. 
We will show that such a solution provides a stable vacuum configuration in a pure
$SU(2)$ QCD.

Let us consider a classical Hamiltonian written in terms of the field $\psi(r,t)$
\bea
H&=&\!\int\!{\rm d}r\, {\rm d}\theta\, {\rm d}\varphi\, \sin\theta
\Big ((\pro_t \psi)^2+(\pro_r
\psi)^2 + \nn\\
&&
\dfrac{1}{2g^2 r^2}(g^2 \psi^2-1 )^2 \Big )
\equiv 4\pi \!\int\! {\rm d}r\, {\cal E}(r,t),  \label{Hamiltonian}
\eea
where ${\cal E}$ is an effective energy density in one-dimensional space.
One has the following non-vanishing field strength components
\bea
F_{r\theta}^2&=&\pro_r \psi,~~~~~~~F_{r\varphi}^1=-\pro_r \psi \sin\theta, \nn \\
F_{\theta\varphi}^3&=&g^2 (\psi^2 -\dfrac{1}{g^2})\sin \theta, \\
F_{t\theta}^2&=&\pro_t \psi, ~~~~~~~F_{t\varphi}^1=-\pro_t\psi \sin\theta,
\nn
\eea
where the radial component of the field strength
$F_{\theta \varphi}^3$ describes spherically symmetric monopole
configuration with a non-vanishing color magnetic flux through a sphere
with a center at the origin $r=0$ \cite{p1}. A color magnetic charge of the
monopole depends on time and radius of the sphere.

One can find an asymptotic
behavior of the stationary solution which
 approaches a standing spherical wave
in the leading order of the Fourier series expansion
\bea
\psi(r,t)&\simeq
& a_0+A_0 \cos (Mr) \sin(Mt)+{\mathcal O}\left(\dfrac{1}{r}\right),
\quad   
\label{asymsol}
\eea
where $a_0$ and $A_0$ are parameters characterizing the mean value and amplitude 
of the standing spherical wave in asymptotic region.
The mass scale parameter $M$ corresponds to the conformal symmetry of the
original Yang-Mills equations.

A local solution near the origin $r=0$ is given by the Taylor series expansion
\bea
\psi &=& \dfrac{1}{g}+\sum_{k=1} c_{2k} (t) r^{2k} ,  \label{locsol}
\eea
where all coefficient functions $c_{2k>2}(t)$ are expressed
in terms of one arbitrary function $c_2(t)$ defining the initial conditions.
The presence of the first term $1/g$ indicates
a non-perturbative origin of the solution. One can verify that such a term
regularizes the singularity of the Wu-Yang monopole and provides
a finite energy density. To find a  stationary solution 
one can impose initial conditions by choosing the function $c_2(t)$
in a simplest form, $c_2(t)=\tilde c_0+\tilde  c_{20}\sin (Mt)$. 
We will choose the initial profile function $c_2(t)$ 
in terms of the Jacobi elliptic function, (\ref{phi0}),
\bea
c_2(t)&=& c_0+c_{20} \text{sn} [M t,-1],\label{initfunc}
\eea
where a set of the parameters $c_0,  c_{20}$ and $M$
provides a uniqueness of the general solution within
a consistent Cauchy problem for the differential equation   (\ref{eqpsi}).
The choice of the initial profile function $c_2(t)$, (\ref{initfunc}), provides  
an additional control of the consistence of numeric calculation
to verify that the numeric solution matches the asymptotic solution (\ref{asymsol})
given precisely by the ordinary sine function $\sin(M t)$ (in the leading order
of the Fourier series decomposition).
A subclass of stationary solutions is classified by one independent parameter,
$c_0$  or $c_{20}$. 

A simple dimensional analysis implies that  the energy corresponding to the
the Hamiltonian (\ref{Hamiltonian}) 
is proportional to the scale parameter $M$. Due to this the energy vanishes in the limit $M\rightarrow 0$. 
This might cause some doubts on existence of a solution.
However, one should stress that standard arguments
on existence of solitonic solutions based on the Derrick's theorem
\cite{derr} can not be applied to the case of stationary solutions 
which satisfy a variational principle of extremal value of the classical action, 
not the energy functional. In addition, in the case of a pure Yang-Mills theory the action is 
 invariant under conformal transformations, 
and its first variational derivative with respect to the scale parameter $M$ equals 
zero identically. So the parameter $M$ represents a moduli space parameter of solutions 
related by conformal transformations (dilatations)  $r \rightarrow Mr, t\rightarrow Mt $. 
Without loss of generality one can fix the value of $M$ to an arbitrary number
which determines the unit of the space-time coordinates.

In order to solve the equation  (\ref{eqpsi})
numerically we choose special values for the parameters,
$g=1, \, M=T_0/(2\pi)$ and $c_0=-0.251$.
The parameter $c_{20}$ is fixed by 
the requirement that a numeric solution should
match the asymptotic solution (\ref{asymsol}).
The mean value $a_0$ and amplitude $A_0$
of the oscillating asymptotic solution are extracted from
the numeric solution which is depicted in Fig.1. 
\begin{figure}[htp]
\centering
\includegraphics[width=70mm,height=55mm,bb=0 0 540 358]{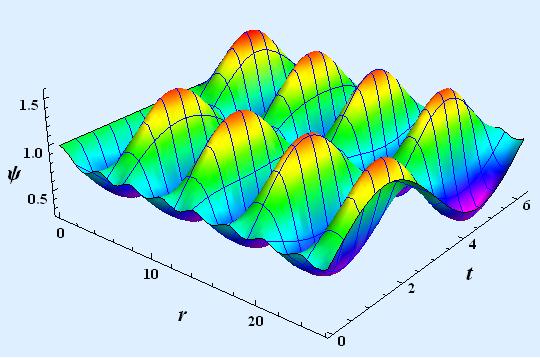}
\caption[fig1]{Stationary spherically symmetric  monopole solution in the numeric domain
$(0\leq r \leq 8 \pi,~0\leq  t \leq 2 \pi)$,~$c_{0}=-0.251$,
$a_0=0.84175$, $A_0=0.6405$.}\label{Fig1}
\end{figure}

Note that at far distance $r_f$ after space-time averaging over the
ring $(r_f\leq r \leq r_f+2 \pi, 0\leq t \leq 2 \pi)$
one gains a partial screening effect for the
monopole charge. The obtained numeric solution
implies an averaged monopole charge at distance $r_f=30$
\bea
g_{\rm m}
&=&\dfrac{1}{4 \pi}\int {\rm d}\theta\, {\rm d}\varphi\, H_{\theta\varphi}
=\dfrac{1}{4 \pi}\int {\rm d}\theta\, {\rm d}\varphi\,
\big(\langle\psi^2\rangle -1 \big) \sin \theta \nn \\
&=& 0.195\cdots,
\eea
with
\bea
\langle\psi^2\rangle&=& \dfrac{1}{4 \pi^2} \int_{r_f}^{{r_f}+2 \pi} \hspace*{-5mm}
{\rm d}r \int_0^{2 \pi}
\hspace*{-3mm}{\rm d}t\, \psi^2(r,t). \nn
\eea
The space-time averaged
magnetic flux of the radial color magnetic field $H_{\theta\varphi}^3$
through a sphere does not vanish in general and depends on the radius of the
sphere.  There is a special non-trivial solution with the parameters
$a_0=0.1\cdots,  \, A_0 = 1.989\cdots$ which corresponds to a totally screened
averaged monopole charge.

With a given numeric monopole solution one can verify the quantum
stability of the monopole field in a similar manner as we 
considered in the previous section.  One should solve the following
Schr\"{o}dinger type eigenvalue equation for possible unstable modes
 (the space indices $(m,n=1,2,3)$ correspond to the
spherical coordinates $(r,\theta,\varphi)$ respectively)
\bea
K_{mn}^{ab} \Psi_n^b(r,\theta,\varphi,t)=\lambda \Psi_m^a(r,\theta,\varphi,t), \label{schr3}
\eea
where $\Psi_n^b(r,\theta,\varphi,t)$ are the wave functions
 describing the quantum gluon fluctuations,
and $K_{mn}^{ab}$ is a differential matrix operator corresponding to one-loop gluon
contribution to the effective action in the temporal gauge
\bea
K_{mn}^{ab}&=&
-\delta^{ab} g_{mn} \pro_t^2
-g_{mn}({\Da}_n {\Da}_n)^{ab} 
+2 \epsilon^{ab}{}_c\Fa_{mn}^c.  \qquad \label{Koper}
\eea

The Schr\"{o}dinger type equation (\ref{schr3}) represents a system
of nine non-linear partial differential equations which should be solved
on three-dimensional numeric domain with sufficiently high numeric accuracy.
An additional technical difficulty in numeric calculation is that one must solve
the equations with changing
the size of the numeric domain in radial direction in the limit $r \rightarrow \infty$
to verify that all eigenvalues remain positive. Fortunately, the numeric analysis
of the solutions corresponding to the lowest eigenvalue is simplified drastically
due to factorization property of the original equation  (\ref{schr3})
and special feature of the class of ground state solutions as we will see below.

The equation (\ref{schr3}) in component form admits factorization,
it can be written as two decoupled systems of partial differential equations
as follows (for brevity of notation we set $g=1$ since the coupling constant
can be absorbed by the monopole function $\psi$)
\begin{widetext}
\bea
(I): &&(\hat\Delta \Psi)_2^2-\dfrac{2}{r^2}\pro_\theta \Psi_1^2+ \dfrac{1}{r^2}\Big ((\psi^2-1)\Psi_2^2-2\psi^2 \Psi_3^1+
2 \csc^2\theta (\Psi_2^2+\Psi_3^1)+2\cot \theta \psi \Psi_3^3
 \Big )
=\lambda \Psi_2^2 , \nn \\
&&(\hat\Delta \Psi)_3^1-\dfrac{2}{r^2}\psi \pro_\theta \Psi_3^3+
\dfrac{1}{r^2}\Big (\psi^2(-2\Psi_2^2+\Psi_3^1)-\Psi_3^1+2\csc^2\theta(\Psi_2^2+\Psi_3^1)+2\cot \theta\Psi_1^2
   \Big ) =\lambda \Psi_3^1 , \nn \\
&&  (\hat\Delta \Psi)_1^2+\dfrac{2}{r^2}\pro_\theta \Psi_2^2
+\dfrac{1}{r^2}\Big ((\cot^2 \theta + \psi^2)\Psi_1^2
+2\cot \theta (\Psi_2^2+\Psi_3^1) +2  \psi \Psi_3^3 +2\Psi_1^2   \Big )
-\dfrac{2}{r} \pro_r \psi \Psi_3^3=\lambda \Psi_1^2 , \nn \\
&& (\hat\Delta \Psi)_3^3+\dfrac{2}{r^2}\psi \pro_\theta \Psi_3^1+
\dfrac{1}{r^2}\Big (2\psi \Psi_1^2+2  \cot \theta \psi (\Psi_2^2+\Psi_3^1)
+2  \psi^2 \Psi_3^3 +\csc^2\theta \Psi_3^3\Big ) -\dfrac{2}{r}\pro_r \psi \Psi_1^2=\lambda \Psi_3^3 ,
\label{eqI}
\eea
\bea
(II):&& (\hat\Delta \Psi)_1^1+\dfrac{2}{r^2}\pro_\theta \Psi_2^1-\dfrac{2}{r^2}\psi \pro_\theta \Psi_1^3
 +\dfrac{1}{r^2}\Big ((2+\cot^2 \theta +\psi^2) \Psi_1^1+2\psi \Psi_2^3
-2\cot \theta (\Psi_3^2-\Psi_2^1) \Big )\nn \\
&&-\dfrac{2}{r}\pro_r \psi \Psi_2^3
=\lambda \Psi_1^1 , \nn \\
&&(\hat\Delta \Psi)_2^3-\dfrac{2}{r^2}\pro_\theta \Psi_1^3+\dfrac{2}{r^2}\psi\pro_\theta \Psi_2^1
+\dfrac{1}{r^2}\Big (2\psi \Psi_1^1
+2\cot \theta \psi (\Psi_2^1-\Psi_3^2)+(2 \psi^2 +\csc^2\theta) \Psi_2^3  \Big )\nn \\
&& -\dfrac{2}{r}\pro_r \psi \Psi_1^1=\lambda \Psi_2^3 , \nn \\
&&(\hat\Delta \Psi)_2^1-\dfrac{2}{r^2}\pro_\theta \Psi_1^1-\dfrac{2}{r^2}\psi\pro_\theta \Psi_2^3
+\dfrac{1}{r^2}\Big (-2\psi\Psi_1^3
+ \psi^2(\Psi_2^1+2 \Psi_3^2) +2\csc^2\theta (\Psi_2^1-\Psi_3^2)-\Psi_2^1  \Big )\nn\\
&&+\dfrac{2}{r}\pro_r \psi \Psi_1^3
=\lambda \Psi_2^1 ,  \nn \\
&&(\hat\Delta \Psi)_1^3+\dfrac{2}{r^2}\pro_\theta \Psi_2^3+\dfrac{2}{r^2}\psi\pro_\theta \Psi_1^1
+\dfrac{1}{r^2}\Big ( 2\cot \theta\psi \Psi_1^1+2(1+\psi^2) \Psi_1^3
-2  \psi(\Psi_2^1+\Psi_3^2) +2\cot\theta\Psi_2^3 \Big ) 
\nn \\
&&
+\dfrac{2}{r}\pro_r \psi(\Psi_2^1+ \Psi_3^2)
=\lambda \Psi_1^3 , \nn \\
&&(\hat\Delta \Psi)_3^2
+\dfrac{1}{r^2}\Big (2\cot \theta (\psi \Psi_2^3-\Psi_1^1)
-2\psi\Psi_1^3+\psi^2(2\Psi_2^1+\Psi_3^2)-2\csc^2\theta (\Psi_2^1-\Psi_3^2)
- \Psi_3^2  \Big )\nn \\
&&+\dfrac{2}{r}\pro_r \psi \Psi_1^3=\lambda \Psi_3^2 ,
\label{eqII}
\eea
where
\bea
\hat \Delta \Psi_m^a \equiv -(\pro^2_{t}+\pro^2_r+\dfrac{2}{r}\pro_r+\dfrac{1}{r^2}\pro^2_\theta
+\dfrac{\cot\theta}{r^2}\pro_\theta)\Psi_m^a.\nn
\eea
\end{widetext}

To solve numerically the systems of equations (I), (II), 
 we choose a rectangular three-dimensional domain
$(0\leq t \leq 2 \pi, r_0\leq r \leq L, 0\leq \theta \leq \pi )$
and use a simple interpolating function for the monopole solution $\psi(r,t)$
\bea
\psi^{\rm int}&=&1-\dfrac{(1-a_0)r^2}{1 + r^2} \nn \\
   && +A_0 (1 -{\rm e}^{-d_0 r^2}) \cos (M r+b_0) \sin(M t), \quad \label{interpolfun}
\eea
where $d_0$ and $b_0$ are fitting parameters.
An obtained numeric solution to the system of equations 
(I), $(\ref{eqI})$, 
implies that the lowest eigenvalue is positive,
$\lambda_{\rm I}=0.0531$, and the corresponding eigenfunctions have the following properties:
the functions $\Psi_1^2$ and $\Psi_3^3$ vanish identically, and remaining two  functions
are related by the constraint $\Psi_3^1=-\Psi_2^2$. So that there is only one
independent non-vanishing eigenfunction which can be chosen as $\Psi_2^2$.
An important feature of the solution corresponding to the lowest eigenvalue
is that the eigenfunction $\Psi_2^2$ does 
not depend on the polar angle, Fig.2.
\begin{figure}[h!]
\includegraphics[width=85mm,height=52mm, bb=0 0 657 316]{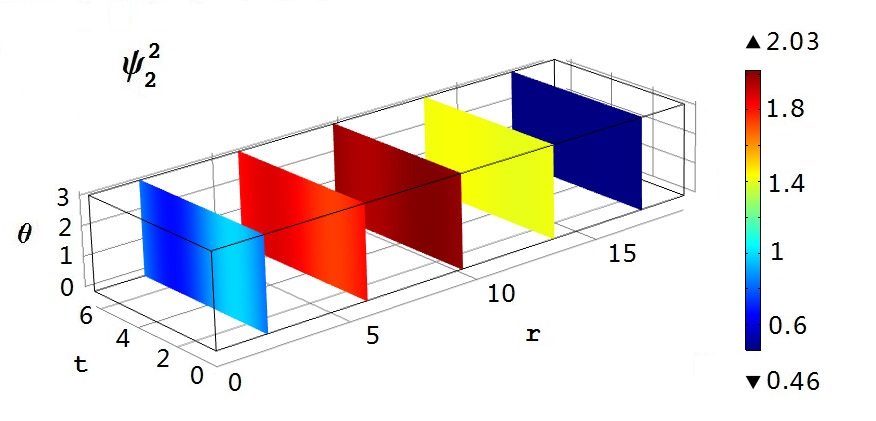}
\caption[fig2]{Eigenfunction $\Psi_2^2$ for the ground state with the lowest eigenvalue
$\lambda_I=0.0531$, $a_0=0.895,~A_0=0.615$, $g=1$, $M=1$, $ 0\leq r \leq 6 \pi$,~ $0\leq t \leq 2 \pi $,
$0\leq\theta \leq \pi$.}\label{Fig2}
\end{figure}
This allows to simplify the system of equations (I)
in the case of solutions corresponding to the lowest eigenvalues.
One can easily verify that system of equations (I), (\ref{eqI}),
reduces to one partial differential equation on two-dimensional space-time
\bea
\Big (-\pro^2_{t}-\pro^2_{r}-\dfrac{2}{r}\pro_r +\dfrac{1}{r^2}(3\psi^2-1) \Big) \Psi_2^2
=\lambda \Psi_2^2.
\eea

The last equation represents a simple Schr\"{o}dinger type equation
for a quantum mechanical problem. The equation does not admit negative eigenvalues if the parameter
$a_0$ of the monopole solution satisfies
the condition $a_0\geq 1/\sqrt 3\simeq 0.577\cdots$
which provides a totally repulsive quantum mechanical potential in this equation.

A structure of the system of equations (II) admits similar factorization properties
on the space of ground state solutions.
We have solved numerically the equations (II), (\ref{eqII}),
with the same background monopole function $\psi(r,t)$ for various values
of the parameters $a_0, A_0, M$. In a special case, $a_0=0.895,~A_0=0.615$, $ 0\leq r \leq 6 \pi$
the obtained numeric solution for the ground state has a lowest
 eigenvalue $\lambda_{\rm II}=0.0142$ which is less than $\lambda_{\rm I}$.
 All components of the solution do not have dependence on the polar angle and
satisfy the following relationships: $\Psi_2^1=\Psi_3^2$ and
$\Psi_1^1=\Psi_2^3=0$. There are two independent non-vanishing functions
which can be chosen as $\Psi_1^3$ and $\Psi_3^2$.
One can check that on the space of solutions corresponding to 
the lowest eigenvalue the system of equations (II), (\ref{eqII}), reduces
to two coupled partial differential equations for two
functions $\Psi_1^3(r,t)$ and $\Psi_3^2(r,t)$
\bea
&&(-\pro^2_{t}-\pro^2_{r}-\dfrac{2}{r}\pro_r ) \Psi_1^3+\dfrac{2}{r^2}\Big( (1+\psi^2)\Psi_1^3-2 \psi \Psi_3^2
\Big ) \nn \\
&&
+\dfrac{4}{r} \pro_r\psi \Psi_3^2=\lambda \Psi_1^3, \nn \\
&& (-\pro^2_{t}-\pro^2_{r}-\dfrac{2}{r}\pro_r ) \Psi_3^2+\dfrac{1}{r^2} \Big ((3 \psi^2-1) \Psi_3^2-2 \psi \Psi_1^3
\Big ) \nn \\
&&
+\dfrac{2}{r}\pro_r \psi \Psi_1^3=\lambda \Psi_3^2. \label{sys2}
\eea
Exact numeric solution profiles for the functions $\Psi_1^3, \Psi_3^2$ are shown in
 Fig.3. 
\begin{widetext}
\begin{figure*}[htp]
  \subfigure[~ $\Psi_1^3$]
 {\includegraphics[scale=0.38,bb=0 0 498 359]{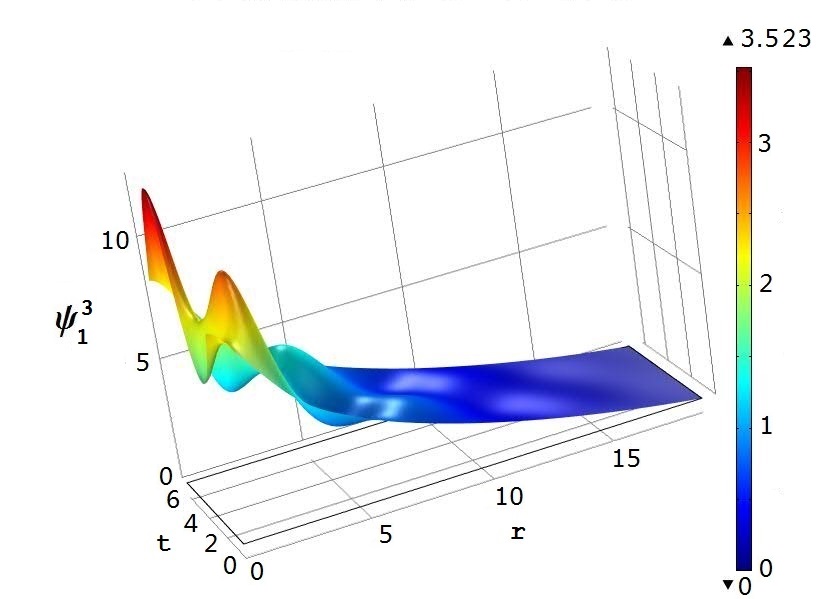}}
\qquad\qquad\qquad
  \subfigure[~ $\Psi_3^2$]
 {\includegraphics[scale=0.38,bb=0 0 498 359]{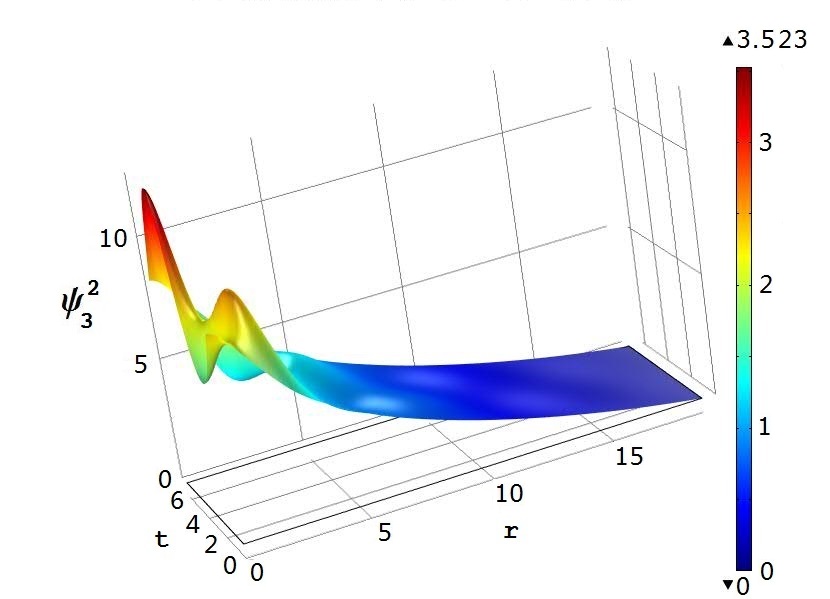}}
\caption[fig3]{Solutions to Eqs.  (\ref{sys2}): the functions
$\Psi_1^3(r,t),~\Psi_3^2(r,t)$ corresponding to the eigenvalue $\lambda =0.014218$,
, $a_0=0.895,~A_0=0.615$, $M=1$,
$ 0\leq r \leq 6 \pi$,~ $0\leq t \leq 2 \pi $: (a)$\Psi_1^3(r,t)$  , (b)  ~$\Psi_2^1(r,t)=\Psi_3^2(r,t)$.}\label{Fig3}
\end{figure*}
\end{widetext}
We have obtained that the lowest eigenvalue is positive
when the asymptotic monopole amplitude $A_0$ is less than
a critical value $a_{\rm 1cr} \simeq 0.625$. 

We conclude, a ground state solution with the lowest eigenvalue 
satisfying the original eigenvalue equation  (\ref{schr3})
can be found by solving a simple system of partial differential
equations (\ref{sys2}).
Note that the numeric solving of the original eigenvalue equations (\ref{schr3})
on a three-dimensional space-time does not provide high enough accuracy,
especially in the case of large radial size of the numeric domain.
This causes difficulty in studying the positiveness of the eigenvalue spectrum
 in the limit of infinite space
when the eigenvalues become very close to zero.
Solving the reduced two-dimensional partial differential equations (\ref{sys2}) 
can be performed easily using standard numeric packages
with a high enough numeric accuracy and convergence.
The obtained numeric accuracy for the eigenvalues $\lambda(L)$
in solving the two-dimensional equations  (\ref{sys2})
is $1.0 \times 10^{-5}$ which allows to construct the
eigenvalue dependence on the radial size $L$ of the space-time
domain in the range  $6\pi\leq L \leq 64 \pi$.
We have proved that the lowest eigenvalue $\lambda (L)$ approaches zero
with increasing $L$ from positive values, as it is shown in Fig.4.
\begin{figure}[htp]
\centering
\includegraphics[width=80mm,height=60mm,bb=0 0 324 199]
{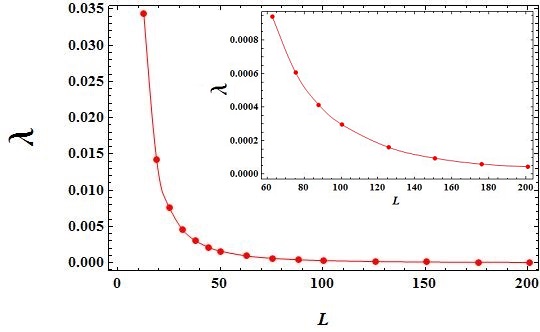}\quad
\caption[fig4]{Lowest eigenvalue dependence, $\lambda(L)$, on the radial size $L$ of the numeric domain.}\label{Fig4}
\end{figure}
This implies that the ground state solution describes the main mode
of the standing spherical wave with a wave vector proportional to the inverse of radial size of the
box, $|\vec p| \simeq 1/L$.  This completes the proof of quantum stability of the
spherically symmetric stationary monopole solution.

The stationary single monopole solution represents a simple example of 
a spherically symmetric vacuum field which has a non-trivial intrinsic microscopic structure
determined by two parameters,
the amplitude $A_0$ and frequency $M$ of space-time oscillations of the monopole field. 
Quantum mechanical consideration implies that
the frequency of vacuum monopole field oscillations has a finite minimal value.
One can estimate a lower bound of $M$ using the condition that 
the characteristic length $\lambda= 2 \pi/M$ of the monopole
field should be less than the hadron size. At macroscopic scale, 
when the observation time is much larger than the period of oscillations of the stationary monopole
solution, the vacuum averaged value of the gauge potential,
$ \langle 0|A_\mu^a|0\rangle$, vanishes
as it should be in the confinement phase.
Contrary to this, the so-called vacuum gluon (monopole) condensate
$H^2\equiv \langle 0|\vec F_{\mu\nu}^2|0\rangle$
does not vanish after averaging over time, and it has inhomogenious 
distribution inside the hadron. 
Calculation of an exact effective action in the case of 
inhomogeneous background vacuum fields represents unresolved problem. 
In the weak field approximation one can
apply the known expression for the Savvidy
renormalized one-loop effective potential 
\cite{savv,yildiz80,claudson80,adler81,dittrich83,flory83,blau91,reuter97,chopakPRD}
\be
V_{\rm eff}(H)
=\dfrac{1}{4} H^2
+\dfrac{11 g^2(\mu)}{96 \pi^2} H^2\Big(\ln \dfrac{g(\mu) H}{\mu^2}
-\dfrac{3}{2}\, \Big),
\ee
where $g(\mu)$ is a renormalized coupling constant
defined at the subtraction point $\mu^2\simeq \Lambda_{\rm QCD}$
($\alpha_s =g^2(\mu)/(4 \pi)\simeq 1$). 
For qualitative estimates we replace the vacuum spherically symmetric 
monopole field $H^2$ with its mean value $\overline {H^2}$ obtained 
by averaging over the space and time. The potential $V_{\rm eff}(\bar H)$
has a non-trivial minimum corresponding to a negative vacuum energy density
at non-zero value of the averaged monopole field, $\bar H_0 \simeq 0.138 \mu^2$ \cite{chopakPRD}.
The value  $\bar H_0$ is consistent with the frequency and amplitude values
$(M\simeq 1,A_0\leq a_{1cr})$ corresponding to stable stationary monopole field
configurations. 

One should stress that the generation of a non-trivial 
vacuum originates from the magnetic moment interaction
between the vacuum magnetic field and quantum gluon fluctuations. Such an interaction
induces the vacuum energy decrease for sufficiently small values of the vacuum monopole condensate
parameter $\bar H$. In the case of the spherically symmetric monopole solution our numeric analysis
confirms that for large values of parameters $M,A_0$, i.e., for large values of  $\bar H_0$, 
the monopole field obtains quantum instability which prevents the generation of a stable 
monopole condensate. 

\section{Discussion}

We have demonstrated that there is a subclass of stationary spherically symmetric monopole solutions
which possesses quantum stability for restricted values
of the amplitude $A_0$ of the asymptotic monopole solution, (\ref{asymsol}).
Recently it has been found that there is another stable stationary
monopole-antimonopole solution in $SU(2)$ and $SU(3)$ QCD \cite{P4}.
This gives a hope that a true vacuum can be formed through condensation
of such monopoles and/or monopole-antimonopole pairs.

Existence of stable monopole field configurations and possible formation 
of a gauge invariant vacuum monopole condensate may shed light on
the origin of color confinement in QCD and give a partial answer to a simple but puzzling question:
why do we have the spontaneous symmetry breaking in the electroweak theory,
while in QCD the color symmetry is preserved despite the similar gauge
group structure in both theories?
The vanishing vacuum averaged value of the gluon field operator
corresponding to the stationary monopole solution,
$\langle A_m^a\rangle$, testifies that there is no spontaneous  symmetry breaking
in QCD in the confinement phase. 
One can apply the ansatz (\ref{ans2}) to electroweak gauge potentials
corresponding to the group $SU(2) \times U_Y(1)$
of the Weinberg-Salam model to find similar stationary electroweak 
monopole solutions. One considers the Higgs complex doublet $\Phi$ in the unitary  gauge,
and choose a simple Dirac monopole ansatz for the hypermagnetic field $\Ba_\mu$
\renewcommand{\arraystretch}{1}
\bea
&&\Phi=  \left (
\ba{cc}
0 \\
 \rho(r,t)
\ea
\right ) ,\qquad \Ba_\mu=\cos \theta. \label{lasteq}
\eea
\renewcommand{\arraystretch}{1.8}
Direct substitution of the ansatz (\ref{ans2}) and the last equations (\ref{lasteq}) into the
equations of motion of the Weinberg-Salam model results in two equations
for two functions $\psi(r,t),~ \rho(r,t)$
\be
\begin{array}{rcl}
\pro^2_{t}\psi-\pro^2_{r}\psi+\dfrac{1}{2}\psi \rho^2
+\dfrac{1}{r^2} (\psi^2-1)&=&0, \\
\pro_{t}^2\rho-\pro^2_{r}\rho-\dfrac{2}{r^2} \pro_r \rho
+\dfrac{1}{2 r^2} \rho \psi^2 +\kappa  \rho(\rho^2-1)&=&0,
\end{array}
\label{wseqs}
\ee
where $\kappa$ is the coupling constant of the Higgs potential.
In a special case of static field configurations the equations
(\ref{wseqs}) reduce to the ordinary differential
equations describing a known Cho-Maison monopole \cite{chomaison}.
A simple numeric analysis of the equations (\ref{wseqs}) shows that
a non-static generalization of the Cho-Maison monopole exists, however it has
the same singularity at the origin $r=0$. We conclude that there is a principal difference
between the Weinberg-Salam model and QCD:
the absence of a regular monopole solution in the Weinberg-Salam model
implies that there is no generation of a stable monopole condensation like in QCD.
This leads to non-vanishing  vacuum averaged values for
the gauge bosons and, consequently, to the spontaneous symmetry breaking.
Contrary to this, in QCD, in the confinement phase, the mean value
of the monopole field $\langle0|A_m^a|0\rangle$ averaged over the
periodic space-time domain vanishes, so that the color symmetry is exact.

In conclusion, we have demonstrated that a classical stationary 
spherically symmetric monopole solution provides
a stable vacuum field configuration in a pure $SU(2)$ QCD.
Generalization of our results to the case of $SU(3)$
QCD is presented in a separate paper \cite{Pa1}. 
The possibility that a stationary classical solution can be
related to vacuum structure is not much surprising since it was
noticed in the past that color magnetic flux tubes in the ``spaghetti''
vacuum should be vibrating from the quantum mechanical consideration
 \cite{olesen81}.
An unexpected result is that a stationary color monopole solution exists
in a pure QCD without any matter fields,  and it possesses remarkable
features such as the finite energy density, total zero spin and existence of intrinsic
mass scale parameter. This gives a strong indication to generation of a stable
vacuum condensate in QCD.


\acknowledgments
One of authors (DGP) acknowledges Prof. C. M. Bai for warm hospitality during his staying at the 
Chern Institute of Mathematics and E. Tsoy for numerous discussions. The work is supported by:
(YK) Rare Isotope Science Project of Inst. for Basic Sci. funded by Ministry of Science, 
ICT and Future Planning, and National Reserach Foundation of Korea, grant NRF-2013M7A1A1075764; 
(BHL) NRF-2014R1A2A1A01002306 and NRF-2017R1D1A1B03028310; 
(CP) Korea Ministry of Education, Science and Technology, Pohang city, and NRF-2016R1D1A1B03932371;
(DGP) Korean Federation of Science and Technology, Brain Pool Program, and grant OT-$\Phi$2-10.


\appendix*

\section{Variational analysis of quantum stability of the stationary monopole field}


To reveal the origin of stability of our numeric solution
we undertake analytic study of the eigenvalue spectrum of the Schr\"{o}dinger type
equation (\ref{schr3}). Since we are interested only in the lowest eigenvalue solution,
one can solve approximately the equation (\ref{schr3}) by applying variational methods.
Within the framework of the variational approach
one has to minimize the following ``energy'' functional
\bea
{\Ha}=\int{\rm d}r\, {\rm d}\theta\, {\rm d}\varphi\,
{\rm d}t\,   r^2\!\sin \theta\, \Psi_m^a K_{mn}^{ab} \Psi_n^b. \label{funct}
\eea
The structure of the kinetic operator $K_{mn}^{ab}$
and the finiteness condition of the functional ${\Ha}$
allow to fix the singularities along the boundaries
$\theta=0$ and $\theta=\pi$ in the
integral density in (\ref{funct}). We factorize the angle dependence
of the ground state wave functions using
the leading order approximation in Fourier series expansion
for the functions $f_m^a$ as follows
\bea
\Psi_1^3(r,\theta,\varphi,t)=f_1^3(r,t), \label{appr1}
\eea
and for other functions
\bea
\Psi_m^a(r,\theta,\varphi,t)&=&f_m^a(r,t) \sin \theta.  \label{appr2}
\eea
With this one can perform the integration in (\ref{funct})
over the angle variables $(\theta, \varphi)$
and obtain an effective ``energy'' functional
\bea
{\Ha}^{\rm eff}&=&\int {\rm d}r\, {\rm d}t\,  r^2
f_m^a  K_{mn}^{ab} f_n^b \nn \\
 &=&\int {\rm d}r\, {\rm d}t\,  r^2 f_m^a\big [g_{mn}\delta^{ab}\tilde K_0
+V_{mn}^{ab}(r,t) \big ]f_n^b, \nn \\
\label{funct2}
\eea
where $\tilde K_0 = -\pro^2_{t}-\pro^2_{r}-(2/r)\pro_r $,
and $V_{mn}^{ab}$ is an effective potential.
The quadratic form
\bea
\langle f|V|f\rangle \equiv \sum_{m,n,a,b} f_m^a V_{mn}^{ab} f_n^b
\eea
contains terms with radial dependencies proportional
to $(1/r^2)$ and $(1/r)$
which correspond to the centrifugal and Coulomb like potentials, respectively.
In the case of a pure Wu-Yang monopole
it was shown that such a background field leads to the
vacuum instability due to the appearance of the attractive
potential $ (-1/r^2)$
in the respective eigenvalue equation for unstable modes
\cite{pak05}. In our case, in the presence of the stationary monopole
solution, one can verify that due to the structure of the local solution near $r=0$
in equation (\ref{locsol})
the quadratic form containing the terms proportional to
$(1/r^2)$ is positively defined for any smooth fluctuating functions
$f_m^a(r,t)$ satisfying the finiteness condition of the ``energy'' functional.
This provides a non-vanishing positive centrifugal potential
in the corresponding Schr\"{o}dinger equation
which prevents appearance of negative eigenmodes
for a special class of background monopole solutions.

By variation of the functional ${\Ha}^{\rm eff}$ with respect to
functions $f_m^a(r,t)$, one obtains the following
effective Schr\"{o}dinger type equation
\bea
 K_{mn}^{ab} f_n^b(r,t)=\lambda f_m^a(r,t). \label{schreff}
\eea
The obtained system of nine differential equations
is explicitly factorized into four decoupled systems of partial differential equations
\bea
(I): && \tilde K_0 f_2^2+\dfrac{1}{r^2} \big ( (5+2 \psi^2) f_2^2+
(6 -4 \psi^2) f_3^1 \big )=\lambda f_2^2, \nn \\
 &&\tilde K_0 f_3^1+\dfrac{1}{r^2} \big ( (6 -4 \psi^2) f_2^2+
(5 +2 \psi^2) f_3^1 \big) =\lambda f_3^1,  \nn\\
\eea
\bea
 (II): &&\tilde K_0 f_1^2+\dfrac{1}{r^2}(3+\psi^2) f_1^2+2 \psi f_3^3)
-\dfrac{2 \psi'}{r} f_3^3=\lambda f_1^2 , \nn \\
& &\tilde K_0 f_3^3+\dfrac{2}{r^2} ( (1+\psi^2)f_3^2+\psi f_1^2)
-\dfrac{2\psi ' }{x} f_1^2=\lambda f_3^3,\nn\\
\eea
\bea
(III): &&\tilde K_0 f_2^1+\dfrac{1}{r^2}((10+4 \psi^2)f_2^1-(12-8 \psi^2) f_3^2\nn \\
&&-3 \pi \psi f_1^3) +\dfrac{3 \pi \psi'}{r} f_1^3=\lambda f_2^1, \nn \\
& &\tilde K_0 f_1^3+\dfrac{1}{2 r^2}(4(1+\psi^2)f_1^3-\pi \psi (f_2^1+f_3^2)) \nn \\
&&+\dfrac{\pi \psi'}{2 r}(f_2^1+f_3^2)=\lambda f_1^3, \nn \\
 &&\tilde K_0 f_3^2+\dfrac{1}{4r^2}(2(5 f_3^2-6f_2^1)+4\psi^2 (2f_2^1+f_3^2) \nn \\
&&-3 \pi \psi f_1^3)+\dfrac{3 \pi \psi'}{4 r} f_1^3=\lambda f_3^2\, , 
\eea
The remaining system (IV) of two equations for the functions $f_1^1, f_2^3$ is
the same as the system (II)
for the functions $f_1^2,  f_3^3$ with the replacement $f_1^2\rightarrow f_1^1$,
$f_3^3\rightarrow f_2^3$.
The obtained equations represent Schr\"{o}dinger type equations
for a charged particle with a positive centrifugal potential
and oscillating Coulomb potential. It is clear that
solutions $\psi(r,t)$ with small enough parameters $a_0, A_0$
will imply a positive eigenvalue spectrum since
the potential with a small enough depth and asymptotic behavior,
${\mathcal O}(1/r^\alpha)$ and
$(\alpha \leq 1)$,  does not lead to bound states in the case
of space dimension $d\geq 3$.
\begin{figure}[h!]
\centering
\includegraphics[width=60mm,height=40mm,bb=0 0 550 359]
{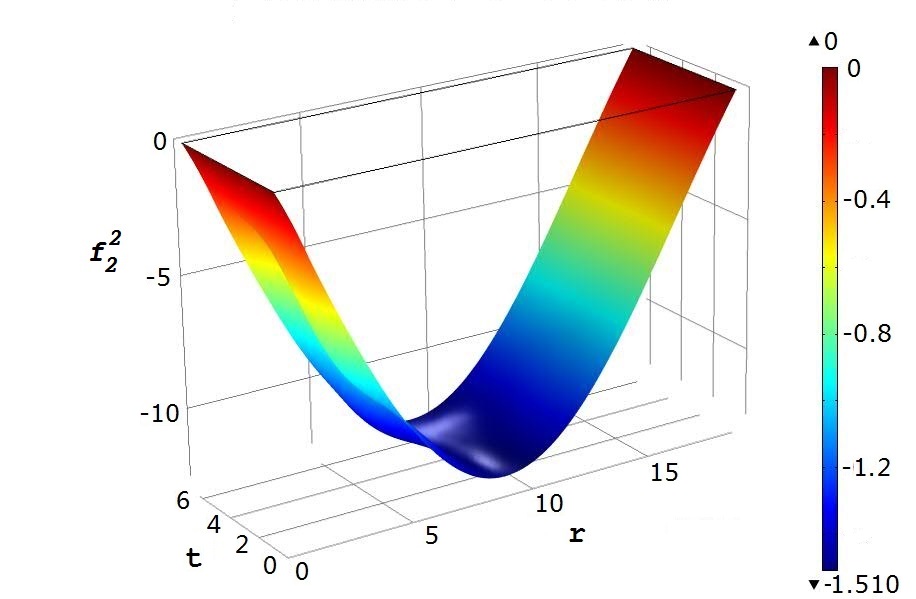}\qquad\qquad\qquad
\caption[fig5]{Solution $f_2^2$ to Eq. (I),~$f_3^1(r,t)=-f_2^2(r,t)$;
$\lambda_I=0.0586$, $ 0\leq r \leq 6 \pi$,~ $0\leq t \leq 2 \pi $.}\label{Fig5}
\end{figure}
\begin{figure}[h!]
\centering
\subfigure[~]
 {\includegraphics[width=40mm,height=40mm,bb=0 0 542 436]
{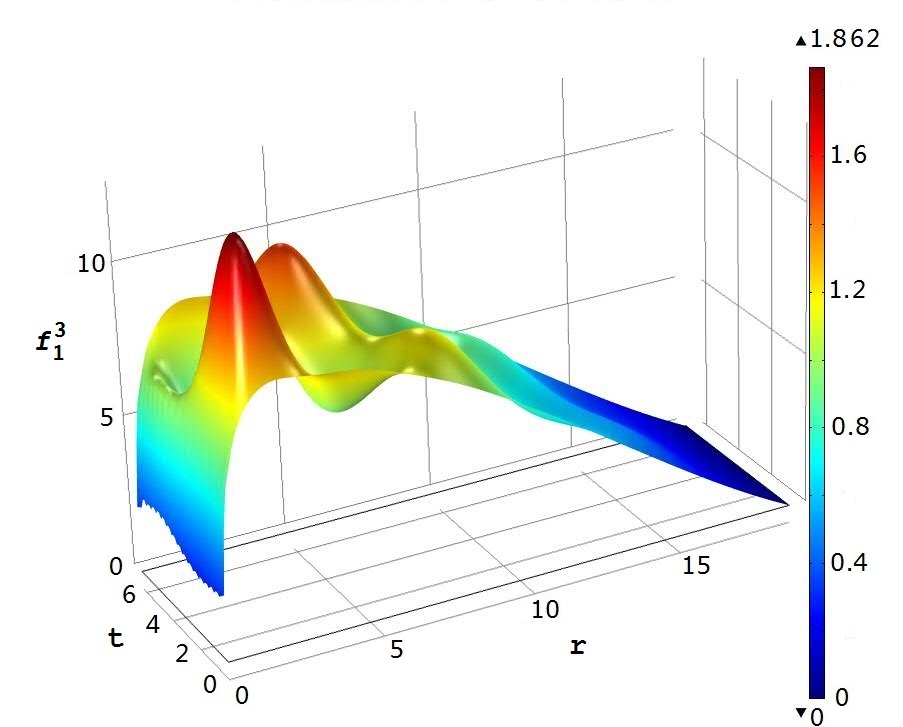}}
\hfill
\subfigure[~]
 {\includegraphics[width=40mm,height=40mm,bb=0 0 542 436]
{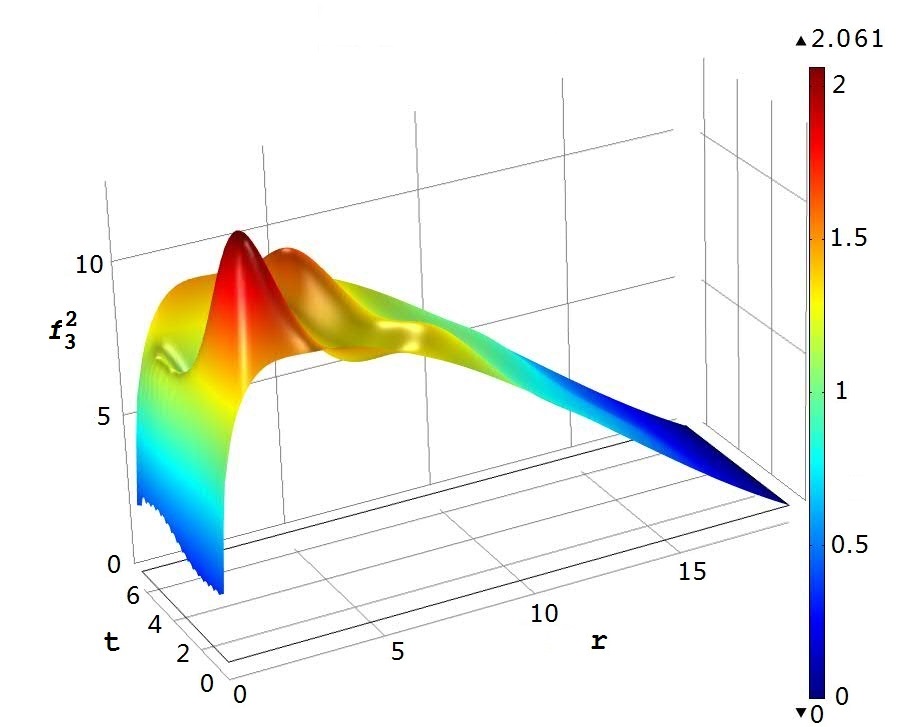}}
\caption[fig6]{Solution to Eq. (III)
$\lambda=0.0293$,
$ 0\leq r \leq 6 \pi$,~ $0\leq t \leq 2 \pi $: (a) $f_1^3(r,t)$, (b) $f_2^1(r,t)=f_3^2(r,t)$.}\label{Fig6}
\end{figure}
Substituting the interpolating function (\ref{interpolfun}) into the Schr\"{o}dinger
equations, one can solve them and obtain the eigenvalue
spectrum. Numeric analysis shows that a complete positive eigenvalue spectrum
exists for solutions $\psi(r,t)$ with parameter values of $a_0$ in a finite range
$0.89\leq a_0 \leq 1$. We solve the equations (I-III)  for a case
of a monopole background field specified by the
parameters $a_0=0.895,~ A_0=0.615$.
A typical profile function for the solutions to equations
(I) and (II) has weak dependence on time, Fig.5.
The corresponding ground state eigenvalues are close to each other,
$\lambda_{\rm I}\simeq 0.0586,~ \lambda_{\rm II}\simeq 0.0552$.
The solution to equation (III)
has a lower eigenvalue $\lambda_{\rm III}\simeq 0.0293$ and manifests
larger time fluctuations as it is shown in Fig.6.
Note that the principal lowest eigenvalue originates from the decoupled system of equations (III)
for the functions $f_2^1, f_1^3$ and $f_3^2$ in qualitative agreement with the results of exact
numerical solving the original eigenvalue equation presented in Section IV. 


\begin{thebibliography}{99}
\bibitem{colorconft} S.J. Brodsky, G.F. de Teramond, and H.G. Dosch,
Int. J. Mod. Phys. {\bf A29}, 1444013 (2014).
\bibitem{nambu74} Y. Nambu, Phys. Rev. {\bf D10}, 4262 (1974).
\bibitem{mandelstam76} S. Mandelstam, Phys. Rep. {\bf 23C}, 245 (1976).
\bibitem{polyakov77} A. Polyakov, Nucl. Phys. {\bf B120}, 429 (1977).
\bibitem{thooft81} G. 't Hooft, Nucl. Phys. {\bf B190}, 455 (1981).
\bibitem{ezawa82} Z. Ezawa and A. Iwazaki, Phys. Rev. {\bf D25}, 2681 (1982).
\bibitem{suzuki80} T. Suzuki, Prog. Theor. Phys. {\bf 80}, 929 (1988).
\bibitem{suganuma95}  H. Suganuma, S. Sasaki, and H. Toki,
Nucl. Phys. {\bf B435}, 207 (1995).
\bibitem{kronfeld87} A. Kronfeld, G. Schierholz, and U. Wiese,
Nucl. Phys. {\bf B293}, 461 (1987).
\bibitem{suzuki90} T. Suzuki and I. Yotsuyanagi,
Phys. Rev. {\bf D42}, 4257 (1990).
\bibitem{stack94} J. Stack, S. Neiman, and R. Wensley, Phys. Rev. {\bf D50},
3399 (1994).
\bibitem{shiba94} H. Shiba and T. Suzuki, Phys. Lett. {\bf B333}, 461 (1994).
\bibitem{bali96} G. Bali, V. Bornyakov, M. M\"uller-Preussker, and
K. Schilling, Phys. Rev. {\bf D54}, 2863 (1996).
\bibitem{savv} G.K. Savvidy, Phys. Lett. {\bf B71}, 133 (1977).
\bibitem{N-O} N.K. Nielsen and P. Olesen, Nucl. Phys. {\bf B144}, 376 (1978).
\bibitem{niel-nino} H.B. Nielsen and M. Ninomiya, Nucl.  Phys. {\bf B156}, 1 (1979).
\bibitem{niel-oles} H.B. Nielsen and P. Olesen, Nucl. Phys. {\bf B160}, 380 (1979).
\bibitem{amb-oles1} J. Ambj{\o}rn and P. Olesen, Nucl.  Phys. {\bf B170}, 60 (1980).
\bibitem{amb-oles2}  J. Ambj{\o}rn and P. Olesen, Nucl.  Phys. {\bf B170}, 265 (1980).
\bibitem{bordag} M. Bordag, Phys. Rev. {\bf D67}, 065001 (2003).
\bibitem{pak05} Y.M. Cho and D.G. Pak,  Phys. Lett. {\bf B632}, 745 (2006).
\bibitem{derr} G.H. Derrick, J. Math. Phys. {\bf 5}, 1252 (1964).
\bibitem{jackiw77} R. Jackiw, The Yang-Mills Vacuum as a Bloch Wave, preprint MIT-CTP-625, (MIT, LNS). Apr 1977. 12 pp.
\bibitem{jackiwRMP} 
R. Jackiw, Rev. Mod. Phys. {\bf 49}, 681 (1977). 
\bibitem{mat1} G.Z. Baseyan, S.G. Matinyan, and G.K. Savvidy, Pisma Zh.
Eksp. Teor. Fiz. {\bf 29}, 641 (1979); JETP Lett. {\bf 29}, 587 (1979).
\bibitem{lahno95} V. Lahno, R. Zhdanov, and W. Fushchych,
J. Nonlinear Math. Phys. {\bf 2}, 51 (1995).
\bibitem{smilga} A. V. Smilga, {\it Lectures on Quantum Chromodynamics},
{\tt [arXiv:hep-ph/9901412]}.
\bibitem{frasca09} M. Frasca, Mod. Phys. Lett. {\bf A24}, 2425 (2009).
\bibitem{tsap} A. Tsapalis, E.P. Politis, X.N. Maintas and F.K. Diakonos,
Phys. Rev. {\bf D93}, 085003 (2016).
\bibitem{p1} B.-H. Lee, Y. Kim, D.G. Pak, T. Tsukioka, and P.M. Zhang,
 Int. J. Mod. Phys. {\bf A32}, 1750062 (2017).
\bibitem{yildiz80} A. Yildiz and P. Cox, Phys. Rev. {\bf D21}, 1095 (1980).
\bibitem{claudson80}  M. Claudson, A. Yilditz, and P. Cox,
	Phys. Rev. {\bf D22}, 2022 (1980).
\bibitem{adler81}  S. Adler, Phys. Rev. {\bf D23}, 2905 (1981).
\bibitem{dittrich83}  W. Dittrich and M. Reuter, Phys. Lett. {\bf B128}, 321, (1983).
\bibitem{flory83}  C. Flory, Phys. Rev. {\bf D28}, 1425 (1983).
\bibitem{blau91}  S.K. Blau, M. Visser, and A. Wipf,
	Int. J. Mod. Phys. {\bf A06}, 5409 (1991).
\bibitem{reuter97}  M. Reuter, M.G. Schmidt, and C. Schubert,
	Ann. Phys. {\bf 259}, 313 (1997).
\bibitem{chopakPRD} Y.M. Cho and D.G. Pak, Phys. Rev. {\bf D65}, 074027 (2002).
\bibitem{schan82} V. Schanbacher, Phys. Rev. {\bf D26}, 489 (1982).
\bibitem{leutwyler} H. Leutwyler, Nucl. Phys. {\bf B179}, 129 (1981).
\bibitem{ragiadakos} C. Ragiadakos,  Phys. Rev. {\bf D26}, 1996 (1982);
Phys. Lett. {\bf B100}, 471 (1981).
\bibitem{parth} R. Parthasarathy, M. Singer, and K.S. Viswanathan,
Can. J. Phys. {\bf 61}, 1442 (1983).
\bibitem{huang} S. Huang and A.R. Levi, Phys. Rev. {\bf D49}, 6849 (1994).
\bibitem{chopakTMU} Y.M. Cho and D.G. Pak,
{\it Dynamical Symmetry Breaking and
Magnetic Confinement in QCD}, Procs. of TMU Symp., Tokyo (2000),
{\tt[arXiv:hep-th/000051]}.
\bibitem{landau9} E.M. Lifshitz and L.P. Pitaevskii,
{\it Statistical Physics, Part 2: Theory of Condensed State}, Vol. 9 (1st ed.),
Butterworth-Heinemann (1980).
\bibitem{p14} M. Luscher, Phys. Lett. {\bf B70}, 321 (1977).
\bibitem{p15} B. Schechter, Phys. Rev. {\bf D16}, 3015 (1977).
\bibitem{p16} H. Arodz, Phys. Rev. {\bf D27}, 1903 (1983).
\bibitem{p17} E. Farhi, V.V. Khoze, and R Singleton, 
Phys. Rev. {\bf D47}, 5551 (1993).
\bibitem{p18} A. Abouelsaood and M.H. Emam, Phys. Lett. {\bf B412}, 328 (1997).
\bibitem{choprd80} Y.M. Cho, Phys. Rev. {\bf D21}, 1080  (1980).
\bibitem{P4}  D.G. Pak, B.-H. Lee, Y. Kim, T. Tsukioka and P.M. Zhang, 
{\tt [arXiv:1703.09635[hep-th]]}. 
\bibitem{chomaison} Y.M. Cho and D. Maison, Phys. Lett. {\bf B391}, 360 (1997).
\bibitem{Pa1}  B.-H. Lee, Y. Kim, D.G. Pak, and T. Tsukioka, \\
{\tt [arXiv:1607.02083[hep-th]]}.
\bibitem{olesen81} P. Olesen, Physica Scripta, {\bf 23}, 1000 (1981).
\end{thebibliography}
\end{document}